\documentclass[iop]{emulateapj}
\usepackage{natbib}
\usepackage{graphicx}
\usepackage{times}
\usepackage{xspace} 
\usepackage{booktabs} 

\slugcomment{\it Accepted Version --- 2017 February 16}
\shorttitle{Extreme Variability in a BAL Quasar}
\shortauthors{Stern et al.}

\def\eg{{e.g.}}
\def\etal{{et al.}}
\def\wise{{\it WISE}}

\def\crts{CRTS~J0841+2005}

\def\deg{\ifmmode {^{\circ}}\else {$^\circ$}\fi}
\def\kms{\ifmmode {\rm\,km\,s^{-1}}\else
    ${\rm\,km\,s^{-1}}$\fi}
\def\ergcm2s{\ifmmode {\rm\,erg\,cm^{-2}\,s^{-1}}\else
    ${\rm\,erg\,cm^{-2}\,s^{-1}}$\fi}
\def\ergAcm2s{\ifmmode {\rm\,erg\,cm^{-2}\,s^{-1}\,\AA^{-1}}\else
    ${\rm\,erg\,cm^{-2}\,s^{-1}\,\AA^{-1}}$\fi}
\def\ergs{\ifmmode {\rm\,erg\,s^{-1}}\else
    ${\rm\,erg\,s^{-1}}$\fi}
\def\kmsMpc{\ifmmode {\rm\,km\,s^{-1}\,Mpc^{-1}}\else
    ${\rm\,km\,s^{-1}\,Mpc^{-1}}$\fi}

\def\civ{\ion{C}{4} $\lambda$1549}

\def\spose#1{\hbox to 0pt{#1\hss}}
\def\simlt{\mathrel{\spose{\lower 3pt\hbox{$\mathchar"218$}}
     \raise 2.0pt\hbox{$\mathchar"13C$}}}
\def\simgt{\mathrel{\spose{\lower 3pt\hbox{$\mathchar"218$}}
     \raise 2.0pt\hbox{$\mathchar"13E$}}}

\def\plotfiddle#1#2#3#4#5#6#7{\centering \leavevmode
\vbox to#2{\rule{0pt}{#2}}
\includegraphics{#1}}

\def\deg{$^{\rm o}$}

\def\arcsec{\ifmmode '' \else $''$\fi}

\def\arcsecpoint{\ifmmode ''\!. \else $''\!.$\fi}

\def\kms{\ifmmode {\rm km\ s}^{-1} \else km s$^{-1}$\fi}
\def\Msun{\ifmmode {\rm M}_{\odot} \else M$_{\odot}$\fi}
\def\Lsun{\ifmmode {\rm L}_{\odot} \else L$_{\odot}$\fi}
\def\Zsun{\ifmmode {\rm Z}_{\odot} \else Z$_{\odot}$\fi}

\def\ergscm2{ergs\,s$^{-1}$\,cm$^{-2}$}
\def\icm3{{\rm cm}^{-3}}
\def\icm2{{\rm cm}^{-2}}
\def\qo{\ifmmode q_{\rm o} \else $q_{\rm o}$\fi}
\def\Ho{\ifmmode H_{\rm o} \else $H_{\rm o}$\fi}
\def\ho{\ifmmode h_{\rm o} \else $h_{\rm o}$\fi}

\def\vFWHM{\ifmmode v_{\mbox{\tiny FWHM}} \else
            $v_{\mbox{\tiny FWHM}}$\fi}
\def\CCF{\ifmmode F_{\it CCF} \else $F_{\it CCF}$\fi}
\def\ACF{\ifmmode F_{\it ACF} \else $F_{\it ACF}$\fi}
\def\Halpha{\ifmmode {\rm H}\alpha \else H$\alpha$\fi}
\def\Hbeta{\ifmmode {\rm H}\beta \else H$\beta$\fi}
\def\Hgamma{\ifmmode {\rm H}\gamma \else H$\gamma$\fi}
\def\Hdelta{\ifmmode {\rm H}\delta \else H$\delta$\fi}
\def\Lya{\ifmmode {\rm Ly}\alpha \else Ly$\alpha$\fi}
\def\Lyb{\ifmmode {\rm Ly}\beta \else Ly$\beta$\fi}
\def\Lyg{\ifmmode {\rm Ly}\beta \else Ly$\gamma$\fi}

\def\ciii{\ifmmode {\rm C}\,{\sc iii} \else C\,{\sc iii}\fi}
\def\civ{\ifmmode {\rm C}\,{\sc iv} \else C\,{\sc iv}\fi}
\def\cv{\ifmmode {\rm C}\,{\sc v} \else C\,{\sc v}\fi}
\def\cvi{\ifmmode {\rm C}\,{\sc vi} \else C\,{\sc vi}\fi}

\def\o5007{[O\,{\sc iii}]\,$\lambda5007$}

\def\o{\o}

\def\gtorder{\mathrel{\raise.3ex\hbox{$>$}\mkern-14mu
             \lower0.6ex\hbox{$\sim$}}}
\def\ltorder{\mathrel{\raise.3ex\hbox{$<$}\mkern-14mu
             \lower0.6ex\hbox{$\sim$}}}
\def\proptwid{\mathrel{\raise.3ex\hbox{$\propto$}\mkern-14mu
             \lower0.6ex\hbox{$\sim$}}}

\begin{document}

\title{Extreme Variability in a Broad Absorption Line Quasar}

\author{Daniel~Stern\altaffilmark{1},
Matthew~J.~Graham\altaffilmark{2},
Nahum~Arav\altaffilmark{3},
S.~G.~Djorgovski\altaffilmark{2},
Carter~Chamberlain\altaffilmark{3},
Aaron~J.~Barth\altaffilmark{4},
Ciro~Donalek\altaffilmark{2},
Andrew~J.~Drake\altaffilmark{2},
Eilat~Glikman\altaffilmark{5},
Hyunsung~D.~Jun\altaffilmark{1},
Ashish~A.~Mahabal\altaffilmark{2},
Charles.~C.~Steidel\altaffilmark{2}
}

\altaffiltext{1}{Jet Propulsion Laboratory, California Institute
of Technology, 4800 Oak Grove Drive, Mail Stop 169-221, Pasadena,
CA 91109, USA [e-mail: {\tt daniel.k.stern@jpl.nasa.gov}]}
 
\altaffiltext{2}{California Institute of Technology, 1200 E.
California Blvd., Pasadena, CA 91125, USA}

\altaffiltext{3}{Department of Physics, Virginia Tech, Blacksburg,
VA 24061, USA}

\altaffiltext{4}{Department of Physics and Astronomy, 4129 Frederick
Reines Hall, University of California, Irvine, CA 92697, USA}

\altaffiltext{5}{Department of Physics, Middlebury College, Middlebury,
VT 05753, USA}

\begin{abstract} 

CRTS~J084133.15+200525.8 is an optically bright quasar at $z =
2.345$ that has shown extreme spectral variability over the past
decade.  Photometrically, the source had a visual magnitude of $V
\sim 17.3$ between 2002 and 2008.  Then, over the following five
years, the source slowly brightened by approximately one magnitude,
to $V \sim 16.2$.  Only $\sim 1$ in 10,000 quasars show such extreme
variability, as quantified by the extreme parameters derived for
this quasar assuming a damped random walk model.  A combination of
archival and newly acquired spectra reveal the source to be an iron
low-ionization broad absorption line (FeLoBAL) quasar with extreme
changes in its absorption spectrum.  Some absorption features
completely disappear over the 9 years of optical spectra, while
other features remain essentially unchanged.  We report the first
definitive redshift for this source, based on the detection of broad
H$\alpha$ in a Keck/MOSFIRE spectrum.  Absorption systems separated
by several 1000 km s$^{-1}$ in velocity show coordinated weakening
in the depths of their troughs as the continuum flux increases.  We
interpret the broad absorption line variability to be due to changes
in photoionization, rather than due to motion of material along our
line of sight.  This source highlights one sort of rare transition
object that astronomy will now be finding through dedicated time-domain
surveys.

\end{abstract}


\section{Introduction}

For galaxies hosting active galactic nuclei (AGNs), time-domain
surveys have long proven to be fertile avenues of research.  Indeed,
optical continuum variability was recognized as a common feature
of quasars shortly after their initial discovery \citep{Matthews:63},
and has since been exploited for purposes ranging from identifying
quasars \citep[\eg,][]{vandenBergh:73}, to determining black hole
masses through reverberation mapping \citep[\eg,][]{Blandford:82,
Bentz:09}, to studying the inner circumnuclear environment
\citep[\eg,][]{Risaliti:02}.  Recent efforts using wide-area,
time-domain surveys have vastly extended this avenue of research
by exploring the optical variability of extremely large samples of
quasars, numbering in the tens to hundreds of thousands
\citep[\eg,][]{MacLeod:12, Graham:14}.  Besides determining the
light curve properties of typical quasars, such work has identified
interesting new phenomenology such as candidate periodic light
curves suggestive of sub-parsec binary super-massive black hole
systems \citep[\eg,][]{DOrazio:15a, DOrazio:15b, Graham:15, Graham:15b,
Jun:15b, Liu:15}, AGN undergoing major flaring suggestive of
microlensing or explosive activity in the accretion disk such as
superluminous supernovae, mergers, or tidal disruption events
\citep[\eg,][Graham \etal, submitted]{Drake:11, Lawrence:16}, and
changing look AGN with the abrupt appearance or disappearance of
broad emission lines \citep[\eg,][]{LaMassa:15, Gezari:17}.

One topic where quasar variability has received particular attention
has been the temporal characteristics of broad absorption line (BAL)
quasars.  Specifically, over the past few years, several teams have
reported on multi-epoch spectroscopic observations of BAL quasars
\citep[\eg,][]{Barlow:92, Lundgren:07, Gibson:08, Gibson:10,
Capellupo:11, Capellupo:12, Capellupo:13, FilizAk:12, FilizAk:13,
FilizAk:14, Vivek:12, He:14, He:15, Joshi:14, Wildy:14, Wildy:15,
Grier:15, Zhang:15}.  While variability in BAL trough strengths is
relatively common, large ($> 50\%$) changes in the absorption
equivalent width is quite rare \citep[\eg,][]{Hall:11}.  A primary
question in BAL variability studies has been whether observed changes
in BAL trough strengths are primarily due to changes in the ionization
state of the outflowing wind \citep[\eg,][]{Wang:15}, or whether
they are due to high column density BAL clouds moving through our
line of sight \citep[\eg,][]{McGraw:15}.




For example, \citet{FilizAk:13} present a detailed analysis of
$\approx 650$ BAL troughs identified in 291 quasars observed by the
Sloan Digital Sky Survey (SDSS), sampling rest-frame timescales
between 1 and 3.7 years.  They estimate that the average lifetime
of a BAL trough is a few thousand years, and that the
emergence/disappearance of BAL features are extremes of general BAL
variability.  \citet{FilizAk:13} also report coordinated BAL
variability across multiple troughs at different velocities.  They
argue that changes in the opacity of the shielding gas producing
changes in the ionizing radiation incident on the BAL material are
the most probable cause for such coordinated variability.


\citet{Grier:15} and \citet{Wildy:15} reach similar conclusions
based on the highly variable BAL lines seen in a spectroscopic
monitoring campaigns.  With variability seen on time-scales of just
a few days, both authors conclude that the most likely cause of
such rapid changes is the BAL gas responding to changes in the
incident ionizing continuum.


Leading to an alternative explanation of BAL variability,
\citet{Capellupo:11, Capellupo:12} report on an ongoing monitoring
campaign of a sample of 24 BAL quasars at $1.2 < z < 2.9$ on
timescales ranging from $\sim 4$~months to $\sim 8$~years.  Studying
the \ion{C}{4} BAL feature, \citet{Capellupo:11} found variability
in 40\% of their sample on month-long timescales, and in 65\% of
their sample on year-long timescales.  They find that higher-velocity
BALs are more likely to vary than lower-velocity BALs, and that
weaker BALs are more likely to vary than stronger BALs.  They suggest
that the observations are best understood as the movement of clouds
within 6~pc of the central engine across the line of sight.  In a
detailed study of the first observation of \ion{P}{5} $\lambda
\lambda 1118, 1128$ BAL variability in a quasar, \citet{Capellupo:14}
argue that the observations are best described by a BAL cloud at a
distance of $\simlt 3.5$~pc moving across the line sight.  The
implied kinetic energy of the outflow would be $\sim 2\%$ of the
quasar bolometric luminosity, which is sufficient to cause substantial
feedback.

Also supporting this interpretation that BAL variability is not
dominated by photoionization, \citet{He:14} report on 18 epochs of
SDSS/BOSS spectroscopy of a BAL quasar at $z = 2.72$.  They find
only a weak correlation between the BAL variability and the continuum
luminosity, suggesting that continuum changes are not driving changes
in the BAL trough amplitudes.

\begin{figure}
\plotfiddle{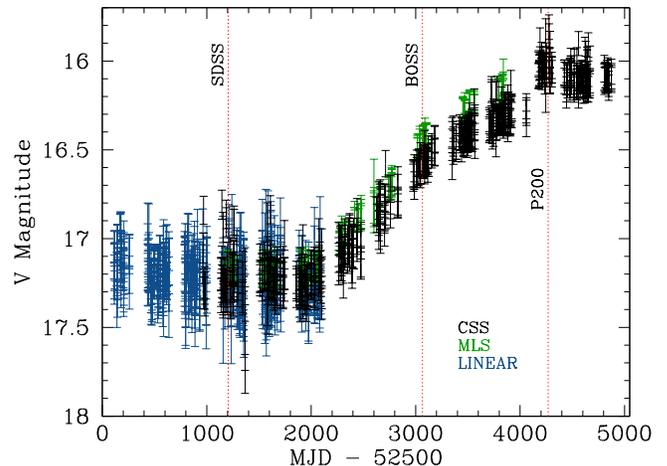}{2.4in}{-90}{32}{32}{-120}{190}
\caption{$V$-band light curve for \crts.  CSS and MLS refer to data
from two of the telescopes that comprise CRTS, the Catalina Sky
Survey 0.7~m Schmidt telescope located at Mt.~Bigelow in Arizona
(CSS) and the 1.5~m Mt.~Lemmon Survey Cassegrain reflector, also
in Arizona (MLS).  LINEAR refers to data from the Lincoln Near-Earth
Asteroid Research program, which used two essentially identical
1.0~m telescopes located at White Sands, New Mexico \citep{Sesar:11}.
To put all data on the same photometric scale, offsets have been
derived from regions of temporal overlap to ensure equal medians.
Vertical lines show the epochs of the optical spectra discussed in
\S2.2.}
\label{fig:lc}
\end{figure}

Here, we report on CRTS~J084133.15+200525.8 (\crts), an optically
bright quasar that has shown extreme variability over the past
decade (Figure~\ref{fig:lc}).  The quasar transitioned from having
a relatively stable visual magnitude of $V \sim 17.3$ between 2002
and 2008, to slowly brightening by a factor of $\sim 2.5$ over the
course of 5 years and then plateauing at $V \sim 16.2$.  As detailed
below, a combination of archival and newly acquired spectroscopy
reveal this source to be an iron low-ionization broad absorption
line (FeLoBAL) quasar exhibiting extreme spectroscopic changes
over the same time period, and the nature of these variations allow
us to assess the likely cause of the BAL trough variability.

Independent of our own work on \crts, \citet{Rafiee:16} recently
reported on this same source as part of a sample of three FeLoBAL
quasars that have shown significant spectroscopic variability over
the past decade.  Interestingly, all three show decreasing strength
of their low-ionization iron absorption.  The current paper
has several additions relative to that work.  Specifically, we
provide new data on \crts, including a new epoch of optical
spectroscopy which demonstrates continued spectral changes, and a
near-infrared spectrum which provides the first precise redshift
for the quasar as well as an estimate of its black hole mass.
Finally, \citet{Rafiee:16} remain agnostic as to whether absorber
transverse motion or ionization variability is the more likely cause
of the changes in the absorption troughs of this source. In contrast,
the additional epoch of Palomar spectroscopy presented here allows
us to argue that ionization variability is the more likely cause
of the extreme absorption variability seen in \crts.


Throughout this paper, we use Vega magnitudes unless otherwise
indicated and we adopt the concordance cosmology, $\Omega_{\rm M}
= 0.3$, $\Omega_\Lambda = 0.7$ and $H_0 = 70\, \kmsMpc$.


%

%
\begin{figure*}
\plotfiddle{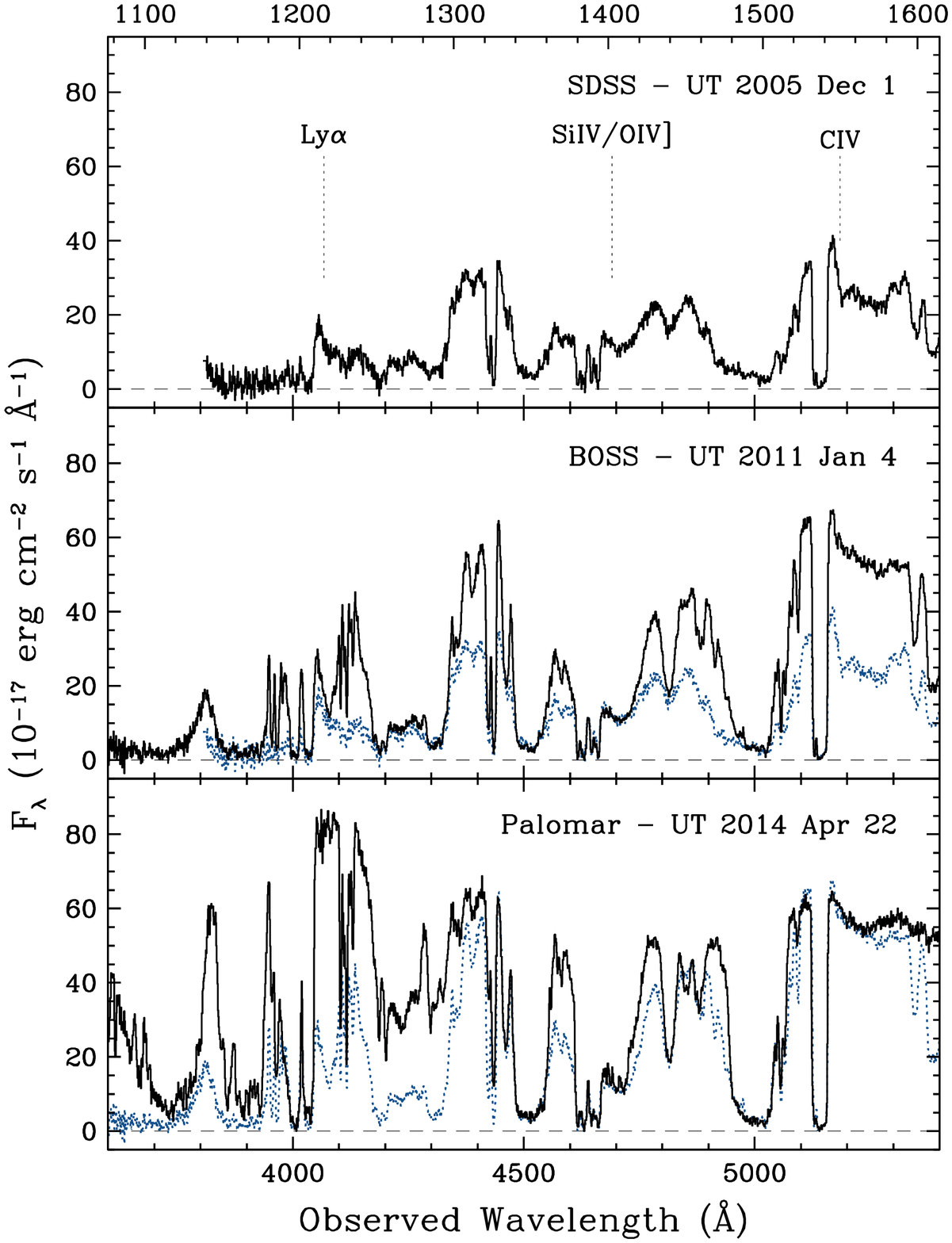}{4.1in}{0}{42}{42}{-240}{-30}
\plotfiddle{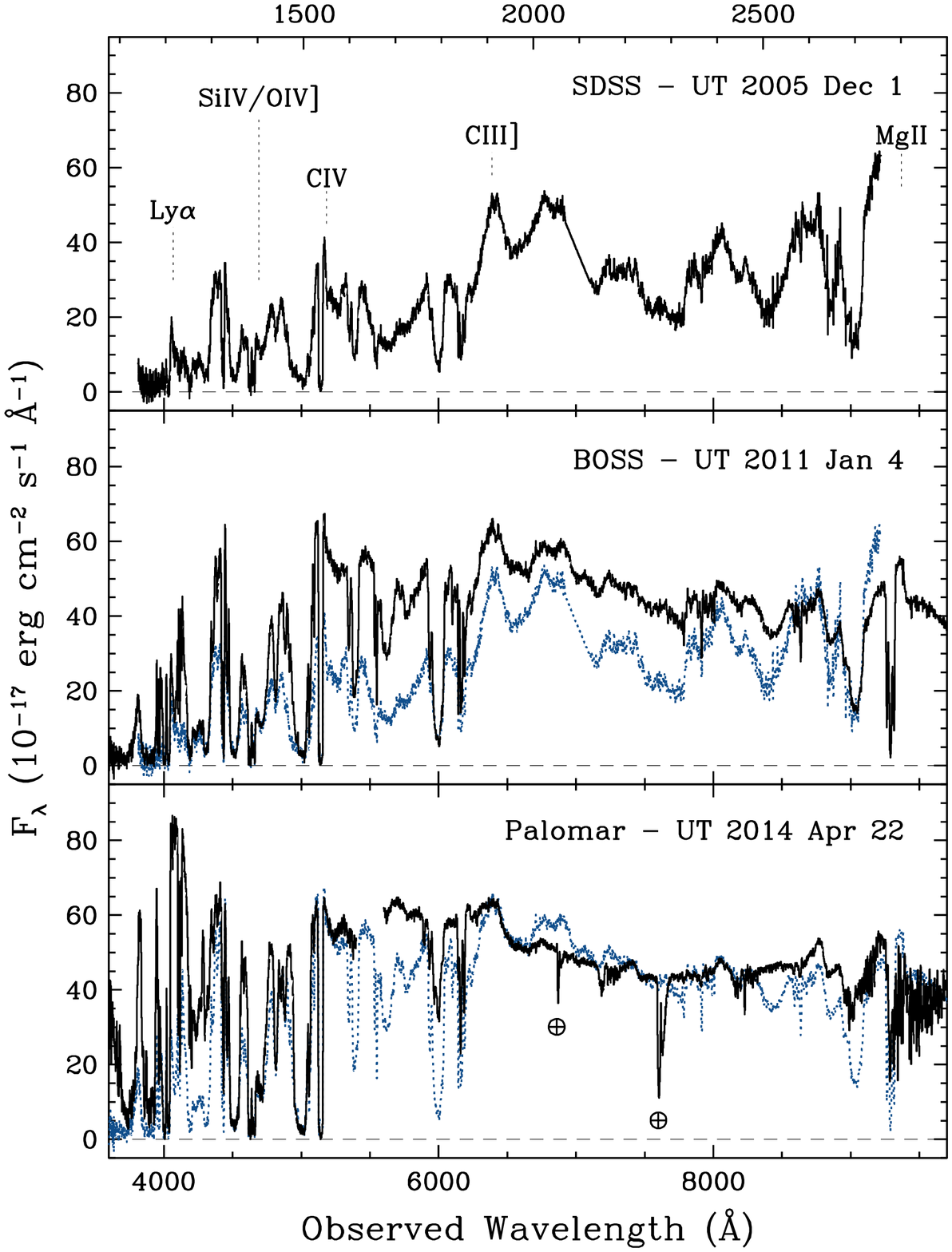}{0.0in}{0}{42}{42}{0}{-13}
\caption{Optical spectra of \crts.  Left panels highlight the bluer
portion of the spectra where many of the strong, variable absorption
features are most evident, while right panels show the full optical
spectra.  Top panels show SDSS spectra obtained in 2005.  Middle
panels show BOSS spectra obtained in 2011 (black solid lines) with
the 2005 SDSS spectra for comparison (blue dotted lines).  Bottom
panels show Palomar spectra obtained in 2014 (black solid lines)
with the 2011 BOSS spectra for comparison (blue dotted lines).
Prominent features are indicated with vertical dotted lines in the
top panels, and the top axis labels refer to the rest-frame wavelength
scale, while the bottom axes refer to the observed frame.}

\label{fig:optical}
\end{figure*}

\section{Data and Results}

\subsection{Optical Light Curve}

The Catalina Real-time Transient Survey\footnote{See {\tt
crts.caltech.edu/}.} \citep[CRTS;][]{Drake:09} leverages the Catalina
Sky Survey, designed to search for near-Earth objects, as a probe
of the time-variable universe.  CRTS has used three telescopes for
much of the past decade, two in the northern hemisphere and one in
Australia, to cover up to $\sim 2500\, {\rm deg}^2$ per night.  The
filterless observations are broadly calibrated to Johnson $V$
\citep[for details, see][]{Drake:13} with a nominal depth of $V
\sim 20$.  The full CRTS data set contains time series for approximately
500 million sources.\footnote{See {\tt http://catalinadata.org/}.}

CRTS represents the best data set currently available with which
to systematically study quasar variability with large samples over
a decade-length timescale.  In an analysis of characteristic
timescales of 240,000 known spectroscopically confirmed objects
using Slepian wavelet variance, Graham et al. (submitted)
originally identified \crts\ as an extreme outlier in the plane
defined by a linear trend (the Thiel-Sen statistic) and deviation
from the median Slepian wavelet variance fit.  In that analysis,
\crts\ has a characteristic timescale $\tau = 109.9$ days, which
is significantly larger than expected for a quasar of its magnitude,
$\tau = 48.0 \pm 5.9$ days.

If we instead characterize quasar light curves with a Gaussian
process damped random walk model and only consider the subset of
79,749 quasars with at least 200 CRTS photometric measurements,
\crts\ again stands out.  The two parameters from this model are
the amplitude, $\sigma$, and the characteristic timescale, $\tau$,
of the damped random walk \citep[\eg,][]{Kelly:09}.  We use a kernel
density estimator to determine the distribution of sources in the
$\sigma-\tau$ plane, and we find that \crts\ resides in an extreme
location in this plane ($\log \Sigma = -7.8$, where $\Sigma$ is the
density of sources in this plane).  Only seven quasars stand out
at this level or more from the population distribution, implying
that only $\sim 1$ in 10,000 quasars show variability behavior as
extreme as \crts.  Further inspection of the CRTS light curve of
\crts\ (Figure~\ref{fig:lc}) also indicates that the variable
behavior is different from the expected stochastic damped random
walk model that describes most quasars, and instead appears more
consistent with a state change.  Further support for this interpretation
comes from earlier photometry of \crts\ reported in \citet{Rafiee:16}
from the Palomar Sky Surveys (POSS-I, Palomar Quick V, and POSS-II),
reaching back to the mid-1950s.  \citet{Rafiee:16} reports no
evidence for a significant change in the optical brightness of
\crts\ prior to 2000.

SDSS imaged \crts\ on UT 2004 December 12 (MJD = 53351), which is prior to the brightening
episode.  The source was unresolved, and based on its unusual and
red colors, SDSS targeted \crts\ for spectroscopic observations as
a high-redshift quasar candidate.

\subsection{Optical Spectroscopy}

\crts\ was first observed spectroscopically by SDSS on UT 2005
December 1 \citep[MJD = 53705;][]{Blanton:03} and was then re-observed
by SDSS-III BOSS on UT 2011 January 4 \citep[MJD = 55565;][]{Dawson:13}.
The spectra, shown in Figure~\ref{fig:optical}, show a source with
many absorption features, making redshift identification challenging.
Indeed, the SDSS data releases have reported a variety of redshifts
for \crts, always with warning flags, ranging from $z = 0.859$ (DR8;
Warning = Many Outliers) to $z = 1.295$ (DR7; zStatus = Failed) to
$z = 3.195$ (DR9; Warning = Negative Emission).  Our visual inspection
of the BOSS spectrum tentatively identified \ion{Mg}{2} and \ion{Fe}{2}
blends in the region around 9400~\AA, implying $z \sim 2.3$,
consistent with both the visual inspection value of $z = 2.342$ in
the SDSS DR12 quasar catalog \citep[DR12Q;][]{Paris:14} and the
results of our Keck infrared spectrum described in \S2.4.



We obtained additional optical spectroscopy of \crts\ using the
Double Spectrograph on the Hale 200'' Telescope at Palomar Observatory
on UT 2014 April 22 (MJD = 56769).  We obtained two 600~s exposures
using the 1.0\arcsec\ slit in cloudy conditions.  The data were
reduced using standard procedures and relative spectrophotometric
calibration was achieved using observations of standard stars
obtained on the same night.  Figure~\ref{fig:optical} presents the
Palomar data, where we have scaled the spectra so that the long
wavelength ($\simgt 5500$ \AA) part of the spectra is of comparable
flux density to the BOSS spectrum at the same wavelengths.


The multi-epoch spectra show the extreme variability exhibited by
\crts, as well as multiple strong absorption features, characteristic
of an FeLoBAL quasar.  FeLoBALs are notoriously challenging targets
for redshift identification \citep[\eg,][]{Becker:97, Brunner:03}.
We see dramatic changes across the full spectrum, particularly in
the spectral region between redshifted Ly$\alpha$ and \ion{C}{4}.
Some features do not change across the near-decade timescale of the
spectroscopy, such as the saturated \ion{C}{4} absorption at 5150
\AA.  Other features completely disappear, such as absorption lines
at $\approx$ 5450 and 8500 \AA.  There is an overall uncovering
of blue continuum emission, with the flux around Ly$\alpha$ increasing
by an order of magnitude over the $>8$~years spanned by the
spectroscopy.  In addition, while the continuum between \ion{C}{3}]
and \ion{Mg}{2} is extremely choppy in the 2005 spectrum, by 2014
it is smoother, which is more typical of normal quasar spectra.

\subsection{Imaging at Other Wavelengths}

\crts\ is a bright near- to mid-infrared source, well detected by
both the Two Micron All Sky Survey \citep[$K_s = 13.62 \pm 0.04$
--- 2MASS;][]{Skrutskie:06} and the {\it Wide-field Infrared Survey
Explorer} \citep[$W3 = 9.58 \pm 0.06$ --- \wise;][]{Wright:10}.
With $W1 - W2 = 0.65$, \crts\ is slightly bluer than the mid-infrared AGN
selection criteria of \citet{Stern:12}, which are $W1 - W2 \geq
0.8$ and $W2 \leq 15.05$.  However, as shown in \citet{Assef:13},
the AGN selection color can be relaxed for brighter sources. 

There is little variability detected at longer wavelengths in this
source.  In AB magnitudes, the $z$-band magnitude recorded by SDSS
was $z = 16.37 \pm 0.01$ on MJD~53351, closely matching the $z$-band
magnitude of $Z = 16.34 \pm 0.01$ recorded by UKIRT Infrared Deep
Sky Survey \citep[UKIDSS;][]{Lawrence:07} on MJD~55141.  In the
near-infrared (in Vega magnitudes), 2MASS recorded $H = 14.41 \pm
0.05$ and $K_s = 13.62 \pm 0.04$ on MJD~51105, closely matching the
UKIDSS values of $H = 14.32 \pm 0.02$ on MJD~54061 and $K = 13.57
\pm 0.02$ on both MJD~54061 and MJD~55238, where we have assumed a
2\%\ floor on the UKIDSS photometric calibration
\citep[\eg,][]{Hodgkin:09}.  Similarly, the mid-infrared flux
measured by {\it WISE} and {\it NEOWISE} \citep{Mainzer:14} varies
by only $\sim 0.04$~mag, comparable to the typical uncertainty.


\crts\ is not detected by {\it ROSAT}, nor was it (serendipitously)
observed by either the {\it Chandra X-Ray Observatory} or {\it
XMM-Newton}.  \crts\ is also not detected by the Faint Images of
the Radio Sky at Twenty cm survey \citep[FIRST;][]{Becker:95},
implying $S_{\rm 1.4~GHz} \simlt 1$~mJy ($5 \sigma$).  Finally, as
expected, observations by the {\it Galaxy Evolution Explorer}
\citep[{\it GALEX};][]{Martin:05}, which sample below the Lyman
limit for $z = 2.35$, do not detect \crts.

\subsection{Near-infrared Spectroscopy}

\begin{figure}
\plotfiddle{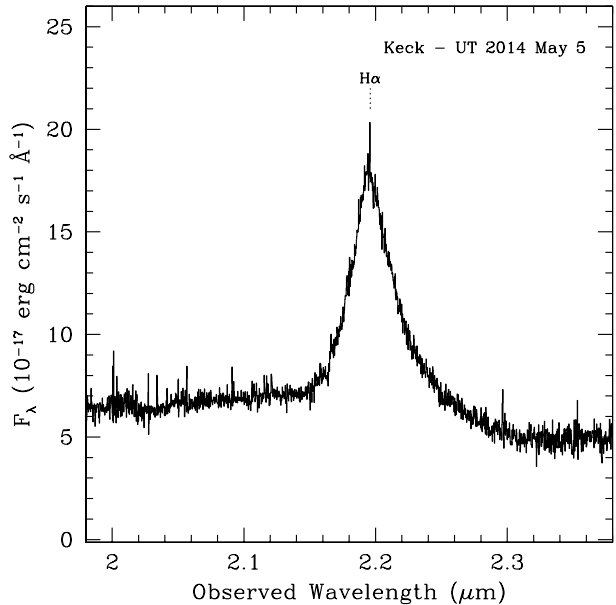}{3.3in}{0}{42}{42}{-120}{-60}
\caption{Near-infrared spectrum of \crts\ obtained with Keck/MOSFIRE.
A single, strong, broad emission line is detected, associated with
H$\alpha$.}
\label{fig:nearIR}
\end{figure}

We obtained a $K$-band (1.95-2.39 $\mu$m) spectrum of \crts\ with
the Multi-Object Spectrometer for InfraRed Exploration
\citep[MOSFIRE;][]{McLean:12, Steidel:14} on UT 2014 May 5 (MJD=56782)
in longslit mode.  We obtained three dithered exposures of 180~s
each through a 0\farcs7 entrance slit under clear conditions with
good seeing.  The spectrum was reduced using a combination of the
MOSFIRE data reduction pipeline (DRP) and custom routines \citep[for
details, see][]{Steidel:14}.  Wavelength calibration was based on
a combination of OH emission lines in the night sky and an internal
Ne arc lamp.  Flux calibration and telluric absorption removal was
accomplished using spectra of an A0V star (Vega analog) observed
at similar airmass.  The final extracted spectrum (Figure~\ref{fig:nearIR})
shows strong continuum and a single broad emission line with a peak
at 2.1956 $\mu$m, which we identify as H$\alpha$ at $z=2.3446$.
The apparent asymmetry in the continuum straddling the line is well
modeled by the \citet{Boroson:92} \ion{Fe}{2} template on the blue
side of the line.

We use the broad H$\alpha$ emission line to estimate the mass of
the black hole in \crts.  First, we apply a multiplicative correction
to the $K$-band spectrum to match the $K$-band photometry from the
UKIDSS observations.  We then approximate the uncertainties in the
spectrum by considering the standard deviation of the spectrum
outside the strong emission line.  We model the H$\alpha$ spectral
region as the sum of two broad Gaussian lines, a single narrow
Gaussian, an iron template (which elevates the continuum on the
blue side of the emission line), and a power-law continuum.  The
full-width at half-maximum (FWHM) of the broad H$\alpha$ emission
is $6086 \pm 42\, \kms$, and the combined luminosity of the broad
H$\alpha$ components is $L_{\rm H\alpha} = (5.56 \pm 0.05) \times
10^{45} \ergs$.  Modeling the broad-band (3000\AA\ to 7$\mu$m)
spectral energy distribution of the quasar as a sum of a power-law
continuum, two blackbody thermal components (500 and 1250~K, to
model the rest-frame IR emission), and line emission from H$\alpha$
and \ion{Fe}{2} as determined from the Keck spectrum, we derive
$L_{5100} = (1.24 \pm 0.02) \times 10^{47} \ergs$.  Following
\citet{Jun:15a}, we derive $\log(M_{\rm BH}/M_\odot) = 10.36 \pm
0.16$ using the $L_{5100}$ estimator and $\log(M_{\rm BH}/M_\odot)
= 10.29 \pm 0.17$ using the $L_{\rm H\alpha}$ estimator.  We note
that these statistical error bars underestimate the true uncertainty,
both due to the non-simultaneity of the imaging and near-infrared
spectroscopy and, more importantly, the systematic uncertainty in
the virial scale factor, $f$, which is the typically the dominant
source of uncertainty in black hole mass measurements; in this case,
we adopt $f = 5.1 \pm 1.3$ from \citet{Woo:13}, as per \citet{Jun:15a}.

For comparison, without access to any well-detected emission features,
\citet{Rafiee:16} simply adopted a black hole mass of $M_{\rm BH}
= 6 \times 10^9\, M_\odot$ as a typical value.  Adopting their value
for the bolometric luminosity of \crts, $L_{\rm bol} = (3.36 \pm
0.69) \times 10^{47}\, {\rm erg}\, {\rm s}^{-1}$ \citep[based on the
observed rest-frame 2900~\AA\ flux density and a bolometric correction
of ${\rm BC}_{2900} = 5 \pm 1$ from][]{Richards:06}, we determine
an Eddington ratio of $L_{\rm bol}/L_{\rm Edd} \sim 0.15$ (in
comparison to their value of 0.45).

%

%
\begin{figure*}
\plotfiddle{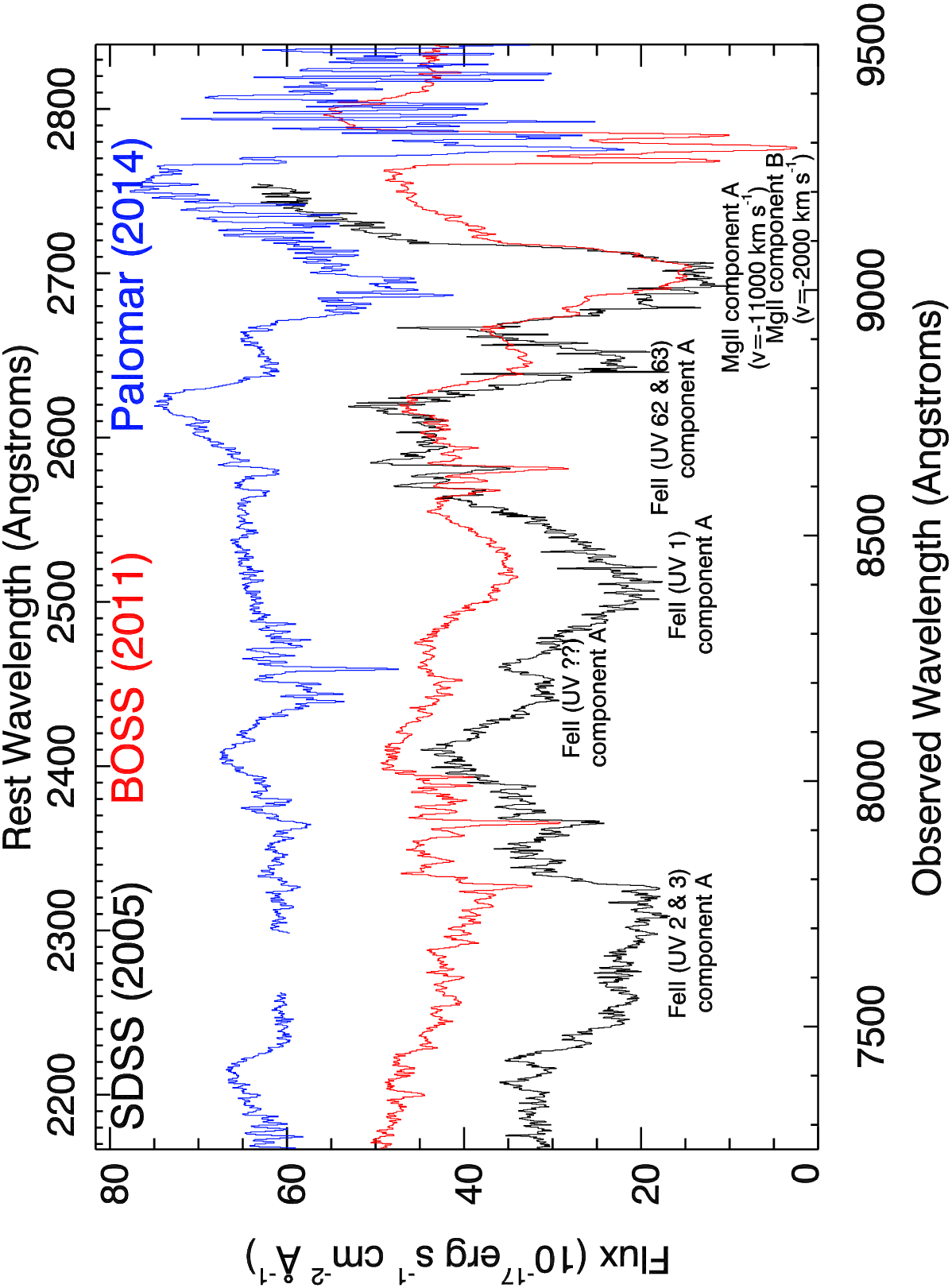}{2.5in}{-90}{32}{32}{-260}{180}
\plotfiddle{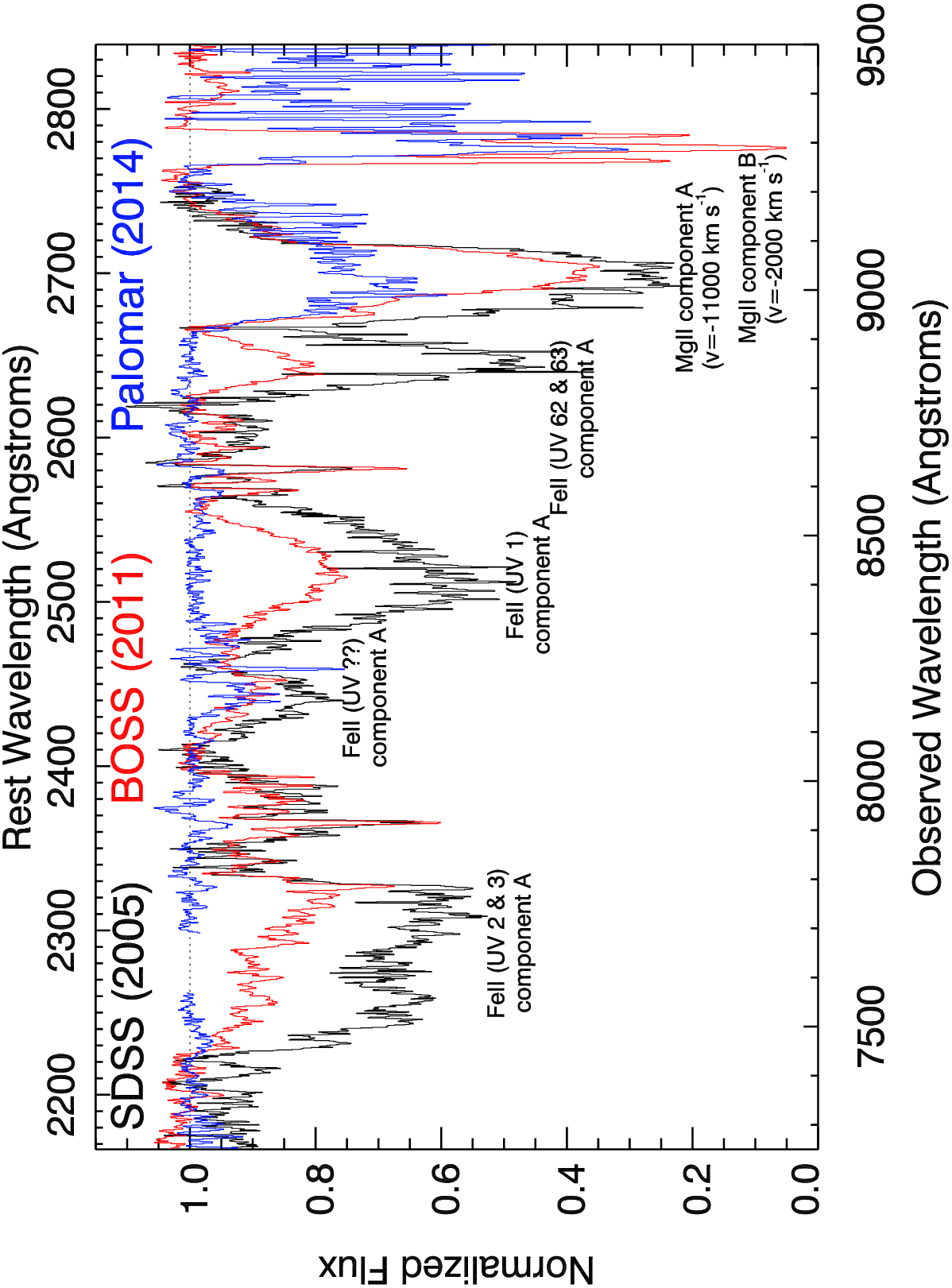}{0.0in}{-90}{32}{32}{0}{198}
\caption{Comparison of the long wavelength portion of all three
spectra with various absorption troughs labeled.  The left panel
shows spectra in units of flux density, while the right panel shows
the same data normalized by a spline-fit to the unabsorbed portions
of the continuum (with the emission lines modeled and divided out).
The right panel makes the coordinated changes in troughs A and B
more evident.  This spectral region covers absorption troughs from
several \ion{Fe}{2} mulitplets as well as the \ion{Mg}{2} doublet.
The depth of all absorption troughs became shallower as the overall
flux level increased with time.}
\label{fig:Fe_Mg}
\end{figure*}

\section{Discussion}


Figure~\ref{fig:Fe_Mg} shows a comparison between the long wavelength
portion of the three epochs of optical spectroscopy.  In the spectral
region beyond $\sim 7300$~\AA\, there is much less blending of
troughs from different ions, making the variability changes simpler
to interpret.  We identify two primary absorption systems.  The
first system, A, shows several troughs of \ion{Fe}{2} UV absorption
between 7500 and 8900~\AA, as well as \ion{Mg}{2} absorption at
9050~\AA.  The other system, B, shows \ion{Mg}{2} absorption at
9300~\AA.  The two absorption systems are separated by 9000~\kms,
yet show coordinated reductions in the depth of their troughs as
the quasar brightens.  This is the expected behavior if the BAL
spectral variability is driven by changes in the photoionization:
as the ionizing continuum flux increases, the column densities of
\ion{Fe}{2} and \ion{Mg}{2} decrease for all clouds along the line
of sight.  (Note that this expectation assumes that the ionizing
continuum changes are correlated with the flux changes around 2500
\AA).  The scenario of clouds moving across our line of sight is
hard pressed to explain both the coordinated changes of the trough
depths as well as the observed trough weakening with increasing UV
flux.  A priori, there is no reason that troughs as widely
separated in velocity as A and B would be correlated since they are
different parcels of gas.  Even more so, there is no reason in this
scenario for changes in the trough depths to be correlated with
flux changes.  Therefore, we interpret the variability in the
absorption troughs to be due to changes in photoionization, rather
than motion of material into our line of sight.  A follow-up paper
will more carefully model the full multi-epoch spectroscopic data
set, including additional spectroscopy from our continuing monitoring,
with the goal of understanding the location and energetics of the
outflow, and its impact on the host galaxy (Chamberlain \etal, in
preparation).

\crts\ appears to be an FeLoBAL quasar in the process of transitioning
to a more common low-ionization BAL (LoBAL) quasar, similar to
FBQS~J1408+3054 reported by \citet{Hall:11}.  We note, however,
that \citet{Hall:11} interpreted the variability in that source as
being related to structure in the BAL outflow moving out of our
line of sight rather than being related to photoionzation changes.

\crts\ highlights the sort of rare, extremely variable quasars that
can be used to probe the physics of quasar outflows.  We expect to
find many more such examples with the new generation of wide-area,
sensitive, high-cadence synoptic surveys.  We were fortuitous in
this case that multi-epoch archival spectroscopy was available for
this source.  In the future, it will be exciting to find similar
major events in real time, allowing real-time multi-wavelength
follow-up in order to more fully dissect the internal workings of
AGN engines.

\acknowledgements 
We thank the anonymous referee for a prompt and helpful referee
report.  CRTS was supported by the NSF grants AST-1313422, AST-1413600,
and AST-1518308.  The work of DS and HJ was carried out at Jet
Propulsion Laboratory, California Institute of Technology, under a
contract with NASA.  DS also acknowledges support from NASA through
ADAP award 12-ADAP12-0109.  NA and CC acknowledge support from NSF
through grant AST~1413319, and from NASA through STScI grants
GO~11686 and GO~12022.  Research by AJB was supported by NSF grant
AST-1412693.  EG acknowledges the generous support of the Cottrell
College Award through the Research Corporation for Science Advancement.
HJ is supported by an appointment to the NASA Postdoctoral Program
at the Jet Propulsion Laboratory, administered by Universities Space
Research Association under contract with NASA.  The authors a
grateful to the staff at the Palomar and Keck observatories, where
some of the data presented here were obtained.  The authors wish
to recognize and acknowledge the very significant cultural role and
reverence that the summit of Mauna Kea has always had within the
indigenous Hawaiian community.  We are most fortunate to have the
opportunity to conduct observations from this mountain.

\smallskip
{\it Facilities:} \facility{CRTS}, \facility{Keck (MOSFIRE)},
\facility{NEOWISE}, \facility{Palomar (DBSP)}, \facility{SDSS},
\facility{WISE}

\smallskip
\copyright 2017.  All rights reserved.

\bibliographystyle{apj.bst}

\clearpage
\end{document}